\documentclass[preprint,sort&compress,12pt]{elsarticle}

\usepackage{graphicx,paralist,url}

\usepackage{amssymb,amsthm,amsmath}

\journal{Physica A}
\begin{document}
\begin{frontmatter}
\title{On interrelations of recurrences and connectivity trends between stock indices}
\author[ipi]{B.~Goswami}
\ead{b.goswami@iiserpune.ac.in}
\author[ipi]{G.~Ambika\corref{cor1}\fnref{fn1}}
\ead{g.ambika@iiserpune.ac.in}
\author[pik]{N.~Marwan}
\ead{marwan@pik-potsdam.de}
\author[pik,ifp]{J.~Kurths}
\ead{juergen.kurths@pik-potsdam.de}
\cortext[cor1]{Corresponding author}
\address[ipi]{Indian Institute of Science Education and Research, Pashan, Pune - 411021, India}
\address[pik]{Potsdam Institute for Climate Impact Research, P.O. Box 60 12 03, 14412 Potsdam, Germany}
\address[ifp]{Department of Physics, Humboldt University Berlin, Newtonstr. 15, 12489 Berlin, Germany}
\fntext[fn1]{+91 20 2590 8037 (Office), +91 20 2589 9790 (Fax)}
\begin{abstract}
Financial data has been extensively studied for correlations using Pearson's cross-correlation coefficient $\rho$ as the point of departure. We employ an estimator based on recurrence plots --- the Correlation of Probability of Recurrence ($CPR$) --- to analyze connections between nine stock indices spread worldwide. We suggest a slight modification of the $CPR$ approach in order to get more robust results. We examine trends in $CPR$ for an approximately 19-month window moved along the time series and compare them to $\rho$. Binning $CPR$ into three levels of connectedness: strong, moderate and weak, we extract the trends in number of connections in each bin over time. We also look at the behavior of $CPR$ during the Dot-Com bubble by shifting the time series to align their peaks. $CPR$ mainly uncovers that the markets move in and out of periods of strong connectivity erratically, instead of moving monotonously towards increasing global connectivity. This is in contrast to $\rho$, which gives a picture of ever increasing correlation. $CPR$ also exhibits that time shifted markets have high connectivity around the Dot-Com bubble of 2000. We stress on the importance of significance testing in interpreting measures applied to field data. $CPR$ is more robust to significance testing. It has the additional advantages of being robust to noise, and reliable for short time series lengths and low frequency of sampling. Further, it is more sensitive to changes than $\rho$ as it captures correlations between the essential dynamics of the underlying systems.
\end{abstract}

\end{frontmatter}
\section{Introduction}
\label{sec1}
Time series of stock markets give some insight into the rather macroscopic dynamics of the underlying systems. An important aspect is to study the interrelations among dynamical stock indices. However, to answer the question: ``How do connections between stock markets change over time?'' a measure of \textit{connectedness} must first be arrived at. In earlier studies, the Pearson's cross-correlation coefficient $\rho$ served as a proxy for `links' between financial data sets \cite{mantegna99,mantegna00,bonanno01,bouchaud01,drozdz01,onnela03,bonanno04,kwapien04,wilcox04,aste05,onnela06,coelho07,coelhohutzler07,cukur07,eom07,tumminello07,conlon09,kenett10}. Its extensive usage has made Pearson's $\rho$ become synonymous with the notion of correlation, and thereby \textit{connections}, itself. However, recent developments in nonlinear data analysis have suggested alternative approaches for estimating connections using, e.g., the study of recurrences, Recurrence Plot (RP), Cross Recurrence Plot (CRP), and Joint Recurrence Plot (JRP) \cite{marwan07}. In our study, we estimate connections between financial data sets from the recurrences of dynamical systems. We use the Correlation of Probability of Recurrence ($CPR$), which is based on RPs and was originally devised to quantify phase synchronization between non-phase-coherent and non-stationary time series \cite{romano05,marwan07}. As the notion of synchronization is innately bound to those of connectedness and co-movement, we argue that $CPR$ too can serve as a measure for connectedness.\\
The potential of recurrences in analyzing financial data has been explored in several studies, e.g., in correlation analysis among currencies \cite{strozzi02}, identification of nature of crashes \cite{guhakurta10}, in estimation of intial time of a bubble \cite{fabretti06} and in quantifying the behaviour of global stock markets during financial crises \cite{bastos11}. Furthermore, cross recurrence analysis has been used to look at synchronicity and convergence of the GDP among member nations of the Euro region \cite{crowley08}.\\
The main objectives of this work are to re-examine commonly held notions of connectedness and explore the potential of using (a slightly modified) $CPR$ as a measure of connectivity between stock indices. We use $CPR$ to formulate connectivity trends between stock indices over almost two decades and compare it to the trends given by $\rho$. We also underscore the importance of using significance tests based on surrogate data sets while interpreting results obtained from field data. In particular, we apply Twin Surrogates \cite{thiel06}, another recurrence-based algorithm, for generating surrogates of our financial time series.\\
We argue for $CPR$ as a suitable measure for measuring connectivity, based on its fundamental nature, and its ability to extract information from relatively poor data sets.\\
In Sec.~\ref{measconn} we outline the basic idea behind our study. Sec.~\ref{theory} outlines the  underlying theory and Sec.~\ref{methods} elaborates on the methods used to analyze the data. Sec.~\ref{disc} states the main results and their implications.
\section{Measuring connectivity}
\label{measconn}
The common thread in connectivity studies of financial data has been \textit{co-movement}, and it has been referred to with varied terms like correlation \cite{mantegna99,mantegna00}, synchronization \cite{imbs04} and cointegration \cite{granger81,granger87}. Each of them incorporates a mathematical formulation that captures some aspect of co-movement for a time series pair. In measuring connectivity we look at another feature that can imply connectedness: \textit{similarity}. If two data sets are \textit{similar} in an aspect that we can measure, then they are closer to each other than, say, with a third data set with which neither of them share as much common ground. Similarity is a more general feature, one of whose manifestations might be co-movement. In our analysis based on a recurrence approach, we intend to take into account the real evolution typical for financial markets better than the classic correlation analysis. Therefore, we make the following basic assumptions:
\begin{enumerate}[i.]
	\item The time series we deal with are the output of \textit{black boxes}, i.e., those systems whose dynamics and model equations are unknown.
	\item The dynamics of such systems may change over time (non-stationarity).
	\item The change in dynamical nature is itself a characteristic feature of the system.
	\item The time series may have features that are common to all of them (e.g. power spectra and clustered volatilities) and further, these similarities are quantifiable.
	\item The quantifiable features are representative of the underlying dynamical nature of the system.
\end{enumerate}
The \textit{quantifiable feature} we study here is the probability of recurrence (see Sec.~\ref{taurec}) and the \textit{similarity} is captured by the cross-correlation between the probabilities of recurrence of pairs of time series (see Sec.~\ref{cpr}).
\section{Theory}
\label{theory}
\subsection{Recurrence Plots}
\label{RPs}
\begin{figure}[tbp]
\centering
\includegraphics[width=\linewidth]{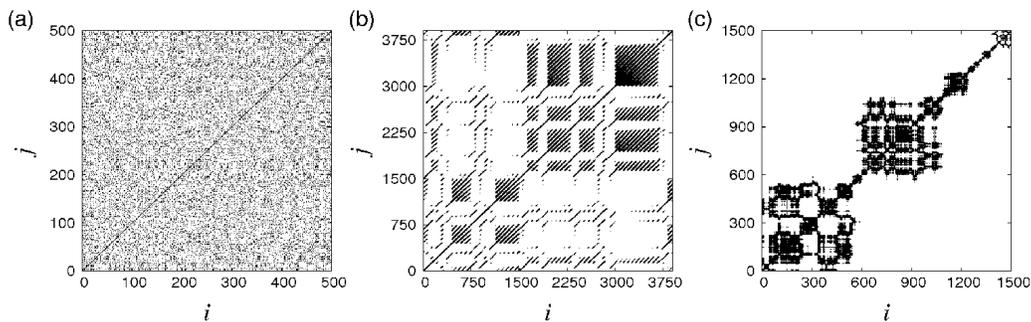}
\caption{\textbf{Recurrence plots for three different types of data}. (a) Uniform white noise. (b) The Lorenz System with $\sigma=10$, $\rho=28$ and $\beta=10/3$. (c) Daily data from DAX. Here, $i$ and $j$ are the index values for the time series entries.}
\label{fig1}
\end{figure}

A \textit{Recurrence Plot} (RP) is a visual tool that shows the recurrence patterns of a dynamical system \cite{eckmann87}. A \textit{recurrence} is defined as the return of the trajectory of a system to an earlier state. In practice, a recurrence is said to occur when the system returns to the neighborhood of an earlier point in the phase space. Mathematically, given a point $\vec{x}_{i} \in \mathbb{R}^{m}$ of a trajectory $\vec{x}_{1}, \vec{x}_{2},\dots, \vec{x}_{N}$, the recurrence matrix $\mathbf{R}$ is estimated as:
\begin{align}
\mathbf{R}_{i,j}(\varepsilon) = \Theta (\varepsilon - \parallel \vec{x}_{i} - \vec{x}_{j} \parallel), ~~~i,j = 1,\dots,N,
\label{RPeq}
\end{align}
where $N$ is the number of points, $\varepsilon$ is an appropriate threshold distance, $\Theta (\cdot)$ is the Heaviside function (i.e., $\Theta (a) = 0$ if $a < 0$, and $1$ if $a \geq 0$) and $\parallel \cdot \parallel$ is an appropriate norm. $\mathbf{R}$ is a matrix of 0s and 1s and an RP is a graphical representation of $\mathbf{R}$ obtained by, e.g., marking a black dot for every 1 and a white dot for every 0.\\
RPs capture the essential features of a system \cite{eckmann87,marwan07}. RPs of three different types of data sets, viz., uniform white noise, chaotic Lorenz system, and daily financial data, are shown in Fig.~\ref{fig1}. All three plots are distinct from each other and characteristic of the system.\\ Next, we give some measures to characterize RPs.
\subsection{Probability of Recurrence}
\label{taurec}
\begin{figure}[tbp]
\centering
\includegraphics[width=\linewidth]{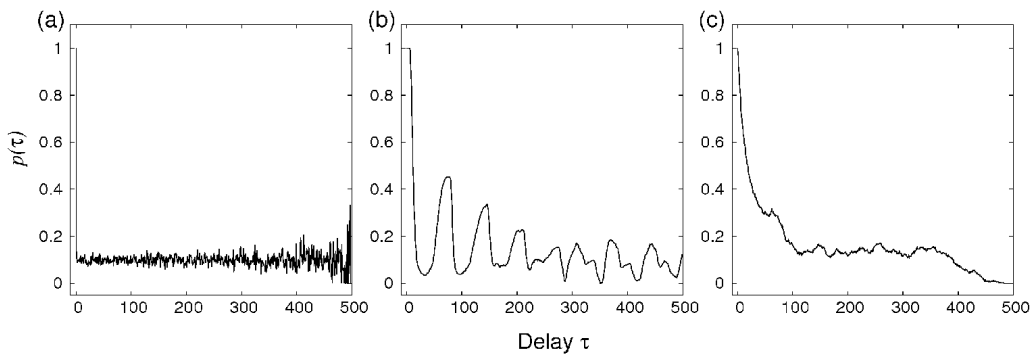}
\caption{\textbf{$p(\tau)$ curves for the three different types of data given in Fig.~\ref{fig1}}. (a) Uniform white noise. (b) The Lorenz System with $\sigma=10$, $\rho=28$ and $\beta=10/3$. (c) Daily data from DAX.}
\label{fig2}
\end{figure}

The \textit{Probability of Recurrence} $p(\tau)$, also called the \textit{$\tau$-recurrence rate}, is the recurrence rate of a diagonal line situated at $\tau$ steps from the main diagonal, i.e., $\mathbf{R}_{i,i+\tau}\, \forall \, i=1,\dots,N-\tau$ \cite{marwan07}. It is a probabilistic measure that gives the probability of an $(i+\tau)^{\textrm{th}}$ point falling in the $\varepsilon$-neighborhood of the $i^{\textrm{th}}$ point:
\begin{align}
p(\tau) = \frac{1}{N - \tau} \sum_{i=1}^{N-\tau} \mathbf{R}_{i,i+\tau}.
\label{taureceq}
\end{align}
It can be considered as a generalized form of an autocorrelation function that statistically reflects the time scales of the system in which it tends to return to a previous configuration. For instance, the uniform white noise time series of Fig.~\ref{fig1}(a) has almost the same probability of recurring to an earlier state for all values of $\tau$ (Fig.~\ref{fig2}(a)), while the chaotic Lorenz system of Fig.~\ref{fig1}(b) has periodic tendencies for high recurrences but with decreasing intensity (Fig.~\ref{fig2}(b)), and the probabilities of recurrence for the daily DAX data from Fig.~\ref{fig1}(c) decreases (without periodicities) with increase in $\tau$, indicating the chances of a drift in the data set (Fig.~\ref{fig2}(c)).
\subsection{Correlation of Probability of Recurrence}
\label{cpr}
\begin{figure}[tbp]
\centering
\includegraphics[width=\linewidth]{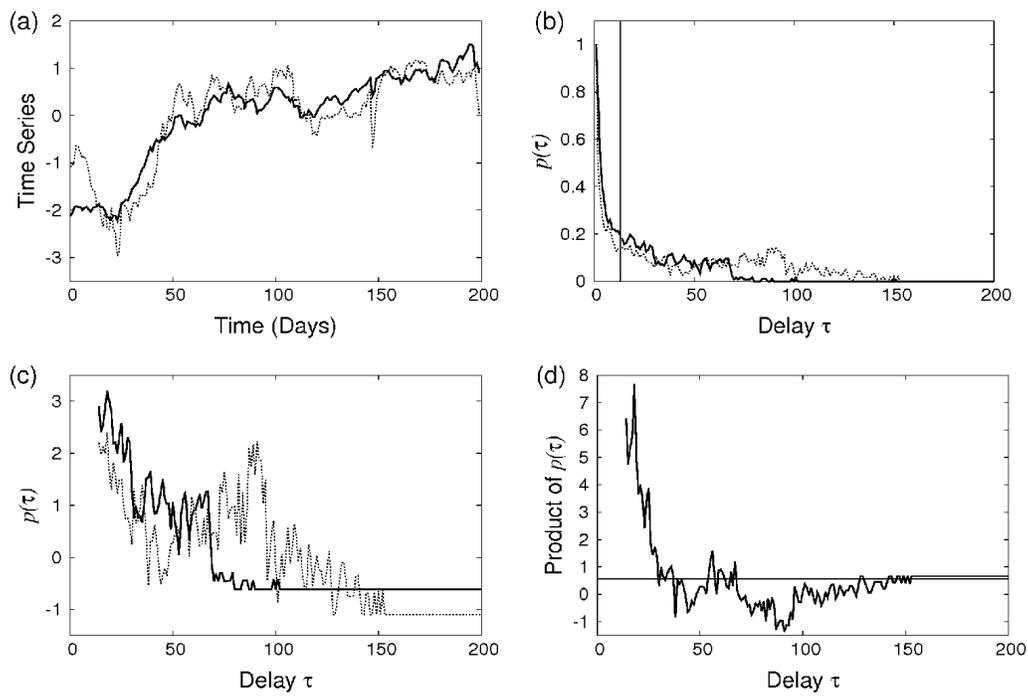}
\caption{\textbf{Calculating $CPR$ from $p(\tau)$}. (a) Normalized daily data for 200 days from two stock indices. (b) $p(\tau)$ curves for these time series. $\tau_{c} = 13$ is shown with a vertical line. (c) Normalized $p(\tau)$ curves beyond $\tau_{c}$ . (d) The product of the two series in (c). The horizontal line is the mean of this series, which is the $CPR$.}
\label{fig3}
\end{figure}

The \textit{Correlation of Probability of Recurrence} ($CPR$) is defined as the cross-correlation coefficient between the probabilities of recurrence of two trajectories $\vec{x}$ and $\vec{y}$ \cite{romano05,marwan07}:
\begin{align}
CPR = \langle \, \bar{p}_{\vec{x}}(\tau) \, \bar{p}_{\vec{y}}(\tau) \, \rangle,
\label{cpreq1}
\end{align}
where $\langle \cdot \rangle$ represents the expectation value and $\bar{x}$ is the series $x$ \textit{normalized} to zero mean and standard deviation one (henceforth, `normalization' refers to this particular way of normalizing a time series).\\
However, all the $p(\tau)$ curves in Fig.~\ref{fig2} start from $p(0) = 1$, because the recurrence rate is always 1 at $\tau = 0$, the main diagonal. This initial portion of the $p(\tau)$ curve, common to all trajectories, introduce a bias towards a high $CPR$ value. To evaluate $CPR$ correctly, we suggest to consider only $p(\tau)$ for such $\tau$ larger than the autocorrelation time $\tau_{c}$ of the system (defined as the delay $\tau$ at which the autocorrelation function of the system  falls to $1/e$):
\begin{align}
CPR = \langle \, \bar{p}_{\vec{x}}(\tau>\tau_{c}) \, \bar{p}_{\vec{y}}(\tau>\tau_{c}) \, \rangle,
\label{cpreq2}
\end{align}
where
\begin{align}
\tau_{c} = max \, \{ \, \tau_{c}(\vec{x}) ,\; \tau_{c}(\vec{y}) \, \}.
\label{cpreq3}
\end{align}
This is shown in Fig.~\ref{fig3} which illustrates the steps involved in estimating $CPR$. The $CPR$ between the two time series according to Eq.~(\ref{cpreq1}) is 0.892, whereas it is 0.575 according to Eqs.~(\ref{cpreq2}) and (\ref{cpreq3}).\\
$CPR$ characterizes the degree of phase synchronization between two time series, with $CPR \approx 1$ implying that the two systems are phase synchronized \cite{marwan07}. However, it can be interpreted more generally as a measure denoting the level to which two trajectories $\vec{x}$ and $\vec{y}$ have similar time scales of recurrence. In this study, this means that two financial time series with a high $CPR$ tend to recur at similar times, suggesting some similarity in their underlying dynamics. 
\subsection{Pearson correlation}
\label{pearson}
The \textit{Pearson correlation coefficient} ($\rho$) is commonly used to analyze correlations between financial data. It is then used to define the \textit{distance} between the data sets as given in \cite{mantegna99}. It is simply the expectation value of the product of the normalized time series:
\begin{align}
\rho_{\vec{x},\vec{y}} = \langle \, \bar{\vec{x}} \, \bar{\vec{y}} \, \rangle,
\label{rhoeq}
\end{align}
where $\langle \cdot \rangle$ and $\bar{x}$ are the same as before.
\subsection{Significance testing and Twin Surrogates}
\label{sigtest}
To use the $CPR$ alone for interpretations might be misleading. In an \textit{active experiment} (such as laboratory experiments or numerical simulations of model systems) where the parameters of the system can be controlled, the $CPR$ for different parameter sets can be compared and thus, a consistent interpretation is possible. However, the financial time series we use are the only realizations of the black boxes generating them. In such a \textit{passive experiment}, where the parameters of the system cannot be changed or controlled, or its dynamics are unknown, it is crucial to generate surrogate time series and check for the statistical significance of the observed $CPR$ against the distribution of $CPR$ obtained from the surrogate data sets. This is done using a statistical test and an appropriate null hypothesis.\\
The \textit{Twin Surrogates} (TS) algorithm is a recurrence-based method of generating surrogate time series \cite{thiel06}. A pair of points $\vec{x_{i}}$ and $\vec{x_{j}}$ of a trajectory $\vec{x}$ (of length $N$) are called \textit{twins} if, for $k = 1,2,\dots, N;\; \mathbf{R}_{k,i} = \mathbf{R}_{k,j}$. This means that, barring their exact positions in the trajectory, these two points have the same neighborhood in phase space. The TS method is an iterative algorithm that involves:
\begin{enumerate}[i.]
\item Identifying twins from the recurrence plot of the trajectory $\vec{x}$.
\item Taking any arbitrary point $\vec{x_{k}} \in \vec{x}$ as the starting point of the surrogate trajectory $\vec{s}$.
\item Iteratively adding subsequent points to $\vec{s}$ as: if $\vec{x_{l}} \in \vec{x}$ is the previous point of $\vec{s}$, and $\vec{x_{l}}$ has no twins, then the next point of $\vec{s}$ is simply $\vec{x_{l+1}}$; whereas if $\vec{x_{l}}$ has $n-1$ twins, then the next point of $\vec{s}$ is any one of that particular set of $n$ twins, chosen with equal probability.
\end{enumerate}
\begin{figure}[hbtp]
\centering
\includegraphics[width=\linewidth]{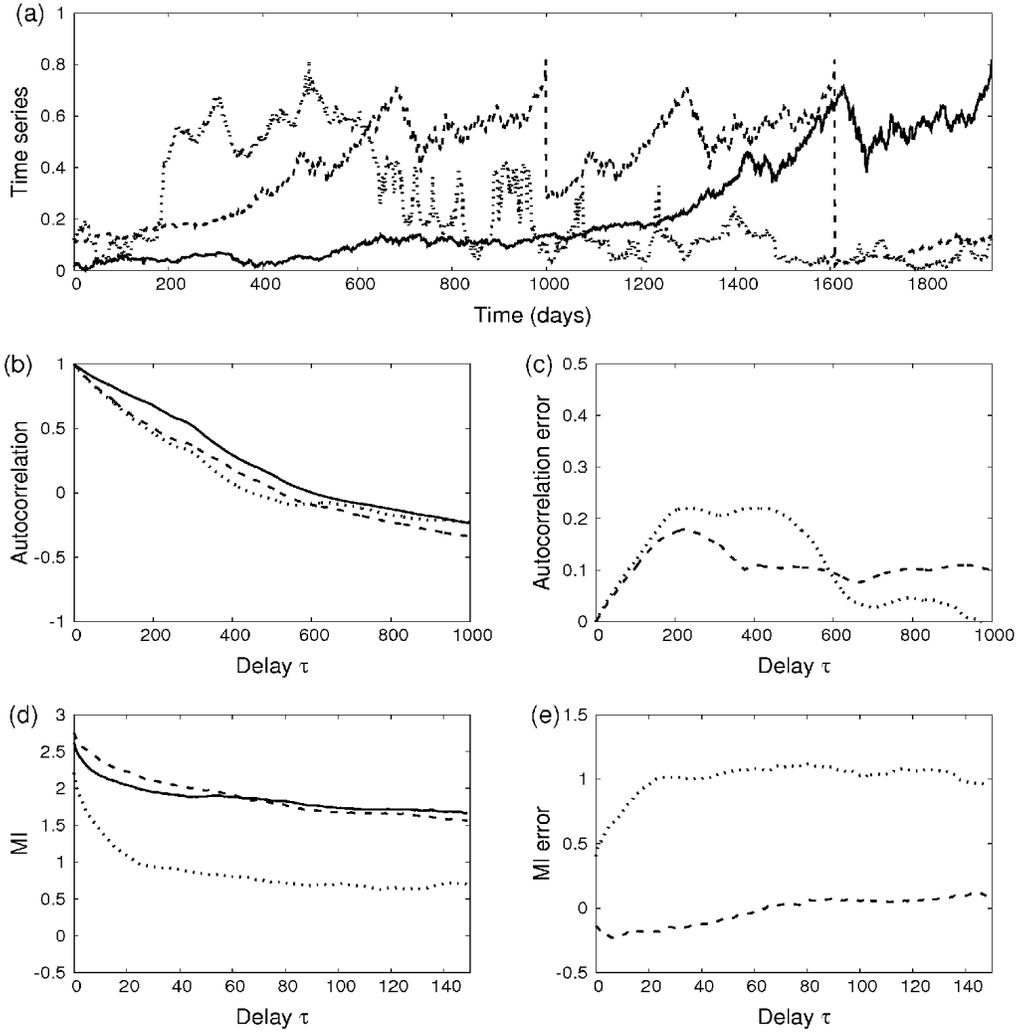}
\caption{\textbf{Comparison of Twin Surrogates and the iAAFT methods}. (a) Daily data for 1942 days from DAX (bold) (scaled to lie between 0 and 1), and a realization of its corresponding Twin Surrogate (dashed) and iAAFT surrogate (dotted). (b) The autocorrelation functions of the three sets of data in (a). (c) The error in autocorrelation of the Twin Surrogate (dashed) and the iAAFT surrogate (dotted). (d) Mutual information of the data sets in (a). (e) The error in mutual information of the Twin Surrogate and iAAFT surrogate. Keys for (c) and (e) are the same as in (b) and (d) respectively.}
\label{fig4}
\end{figure}
Although there are several methods to generate surrogate time series, it is natural to use TS in this study. In the widespread \textit{iterative Amplitude Adjusted Fourier Transform} (iAAFT) method, surrogates are created with the assumed null hypothesis that the observed time series is the result of a Gaussian process seen through a static linear filter whereas, in TS, each surrogate is an independent realization of the observed time series differing only in the initial conditions. iAAFT surrogates do not preserve nonlinear characteristics such as mutual information, whereas TS preserves linear as well as nonlinear properties of a system \cite{thiel06}. Fig.~\ref{fig4}(a) shows the original time series from DAX and both the surrogates via iAAFT and TS. Although the errors in autocorrelation for both methods are comparable, the iAAFT surrogate has much larger error in mutual information than the TS (Figs.~\ref{fig4}(b)-(e)).\\
We use the TS to generate a test distribution using which we test the significance of the observed measure $M$ (which can be either $CPR$ or $\rho$). The null hypothesis for this significance test is that each surrogate is an independent trajectory of the same dynamical system which gave rise to the observed time series. This means that when we test for significance, we check for the probability that an independent realization of one of the time series (as given by its TS) can give a similar value of $M$. The steps involved in the test are:
\begin{enumerate}[i.]
	\item The value of $M$ between time series A and B is estimated and designated as  $M_{o}$ (say).
	\item TS are generated from the series B.
	\item $M$ is calculated between each surrogate and the series A. This gives the test distribution of $M$. We assume that this distribution is roughly normal as each TS corresponds to a distinct trajectory of system B starting from an independent initial condition.
	\item The mean $\mu$ and standard deviation $\sigma$ are estimated for this test distribution.
	\item The test statistic $Z$ is then evaluated as:
	\begin{align}
Z = \frac{M_{o}-\mu}{\sigma},
\label{ztesteq2}
\end{align}
which can be used to infer the probability with which $Z$ belongs to a standard normal distribution.
\end{enumerate}
A prefixed cut-off for the probability is decided below which $M$ is said to be \textit{significantly different} from the test distribution, and this is the \textit{significance level} of the test. The probability value (or $p$\textit{-value}) obtained from the standard normal table for the test statistic $Z$ represents the probability that $M$ might actually be from the test distribution. In terms of the null hypothesis --- that the two time series are independent --- the $p$-value represents the probability that the null hypothesis cannot be rejected.
\section{Methods}
\label{methods}
\subsection{The data set}
\label{dataset}
The daily close values ranging from 3$^{\textrm{rd}}$ December 1990 to 30$^{\textrm{th}}$ April 2010 of nine stock indices (given in Table~\ref{tbl1}) from around the world were used in our analysis. Three were from Asia, three from Europe and three from the U.S.A. The data was obtained from \url{http://finance.yahoo.com/}. Each data set contained a numerical value (representing the close value of the index on a particular date) and its corresponding date.
\begin{table}[tbp]
\begin{center}
\caption{Test data: Market indices and their locations}
\label{tbl1}
\footnotesize
\begin{tabular}{cll}
\\
\hline
Label & Stock Index & Location \\
\hline
 A &  CAC 40 				& France  \\
 B &  FTSE 100 			& U.K.  \\
 C &  DAX 					& Germany  \\
 D &  NASDAQ 				& U.S.A  \\
 E &  DJIA 					& U.S.A  \\
 F &  S\&P 500 			& U.S.A  \\
 G &  Nikkei 225 		& Japan  \\
 H &  Hang Seng 		& China  \\
 I &  Strait Times  & Singapore  \\
\hline
\end{tabular}
\end{center}
\end{table}
\subsection{Preprocessing the data set}
\label{preprocess}
\begin{figure}[tbp]
\centering
\includegraphics[width=\linewidth]{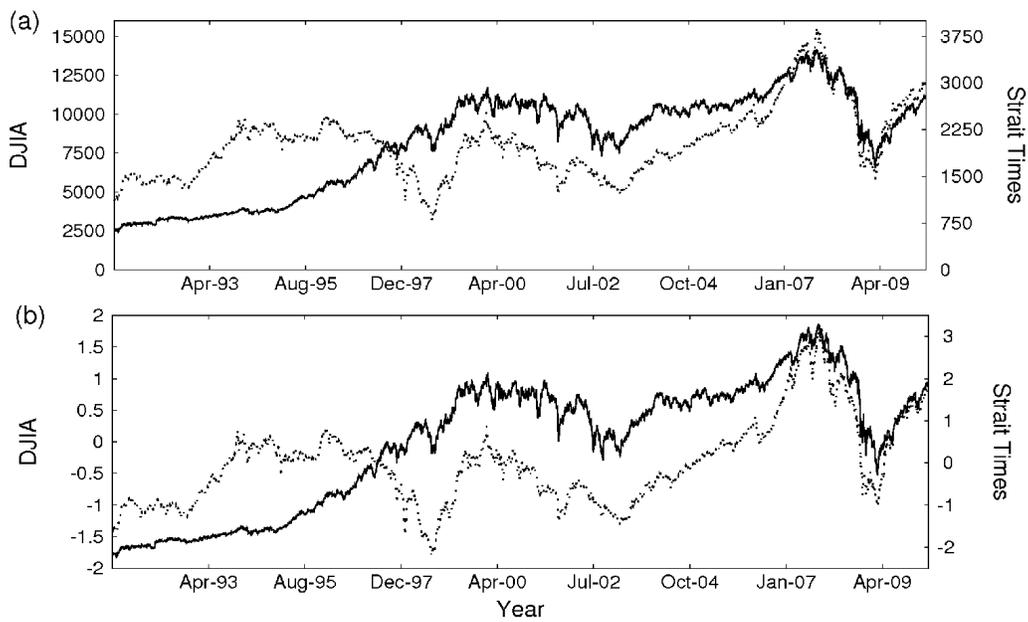}
\caption{\textbf{Normalizing the data}. (a) Daily close values for Dow Jones Industrial Average (DJIA) (bold)and Strait Times (dotted). (b) The time series in (a) after normalization (note the difference in the vertical axes in both figures). The left vertical axis is for DJIA while the right vertical axis is for Strait Times.}
\label{fig5}
\end{figure}

The data sets had unequal lengths because of the unequal distribution of holidays for stock indices in different regions of the world. To align them temporally, mismatched dates (and the corresponding close values) were deleted, i.e., any date of a particular market not present in any one (or more) other market(s) of the remaining eight resulted in the deletion of that date (and the corresponding close value) from all the markets. Simply put, a common intersect of all the nine data sets, in terms of dates, was obtained, which was of length 4238 time points.\\
Moreover, the index values of the different indices were arbitrary (Fig.~\ref{fig5}(a)). To make qualitative comparisons possible, they were \textit{normalized} to mean zero and a standard deviation of one (Fig.~\ref{fig5}(b)). This enabled us to compare the scaled time series and all the recurrence-based measures obtained from them using the same value of the parameters, e.g., the same value of the recurrence threshold $\varepsilon$.
\subsection{Selection of parameters: Preliminary analysis}
\label{prelim}
\subsubsection{Window size}
\label{windowsize}
\begin{figure}[tbp]
\centering
\includegraphics[width=\linewidth]{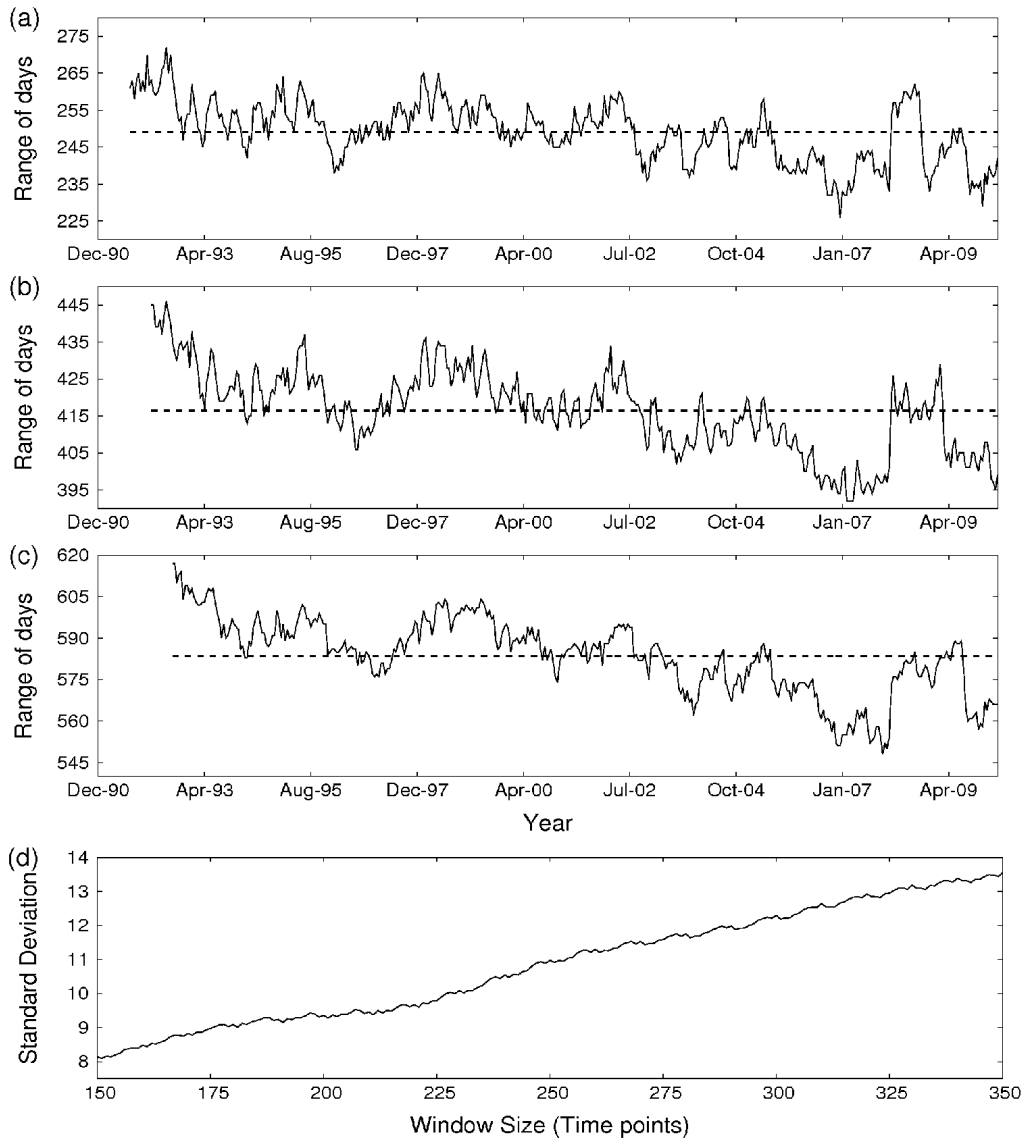}
\caption{\textbf{Selecting the window size}. (a)-(c) The range of days contained in windows of sizes 150, 250 and 350 time points as they are moved along the time series in steps of 10 time points. The horizontal dashed lines represented the mean number of days. (d) The standard deviation of the distribution of range of days contained in windows sized between 150 to 350 time points.}
\label{fig6}
\end{figure}
The primary method in this study is to slide a window of fixed width along the time series and then estimate $CPR$ or $\rho$. However, because of the deletion of dates (see Sec.~\ref{preprocess}), if a window of, say, 500 consecutive time points is chosen, the actual \textit{range} of days contained in it is larger than 500 and further, this range changes as the window moves along the time series (Fig.~\ref{fig6}(a)-(c)). In order to get an understanding of the variation in the range of days contained in a window with respect to the number of time points in the window, we varied the window size from 150 to 350 time points and estimated the standard deviation of the range of days contained in each window, and found that the standard deviation increases (almost) monotonically with increasing window size (Fig.~\ref{fig6}(d)).\\
Ideally, we want a window with negligible standard deviation. However, this would mean reducing the window size which would reduce the effectiveness of the RP and, in turn, the measures estimated from it. As a compromise, we choose a window of 250 time points, which is partly arbitrary but we (reasonably) assume that the qualitative features of the results are not severely effected by increasing or decreasing the window size from 250 by a small margin. The mean range of days contained in a 250 time point window is approximately 416 days, which is roughly equal to 19 months (considering a 5-day week).
\subsubsection{RP parameters}
\label{rpparam}
There are several ways of constructing an RP from data \cite{marwan07}, e.g., with a fixed threshold $\varepsilon$ or a varying one, with or without embedding. In our analysis, we do not embed the time series. Also, there is no fixed rule to select $\varepsilon$. It depends on the objectives of the study and the nature of the data set. The choice of $\varepsilon$ should ensure that the recurrence matrix $\mathbf{R}$ represents the dynamics of the system. It should neither be too large (to avoid counting spurious recurrences) nor be too small (to avoid excluding crucial recurrences). One rule of thumb is to ensure that $\varepsilon$ does not exceed 10\% of the maximum distance between the points in the time series \cite{zbilut92}. Another approach is to consider the recurrence-based measure as a signal detector and then choose the $\varepsilon$ value that yields the maximum power from the signal for the detector being considered \cite{schinkel08}.\\
\begin{figure}[btp]
\centering
\includegraphics[width=\linewidth]{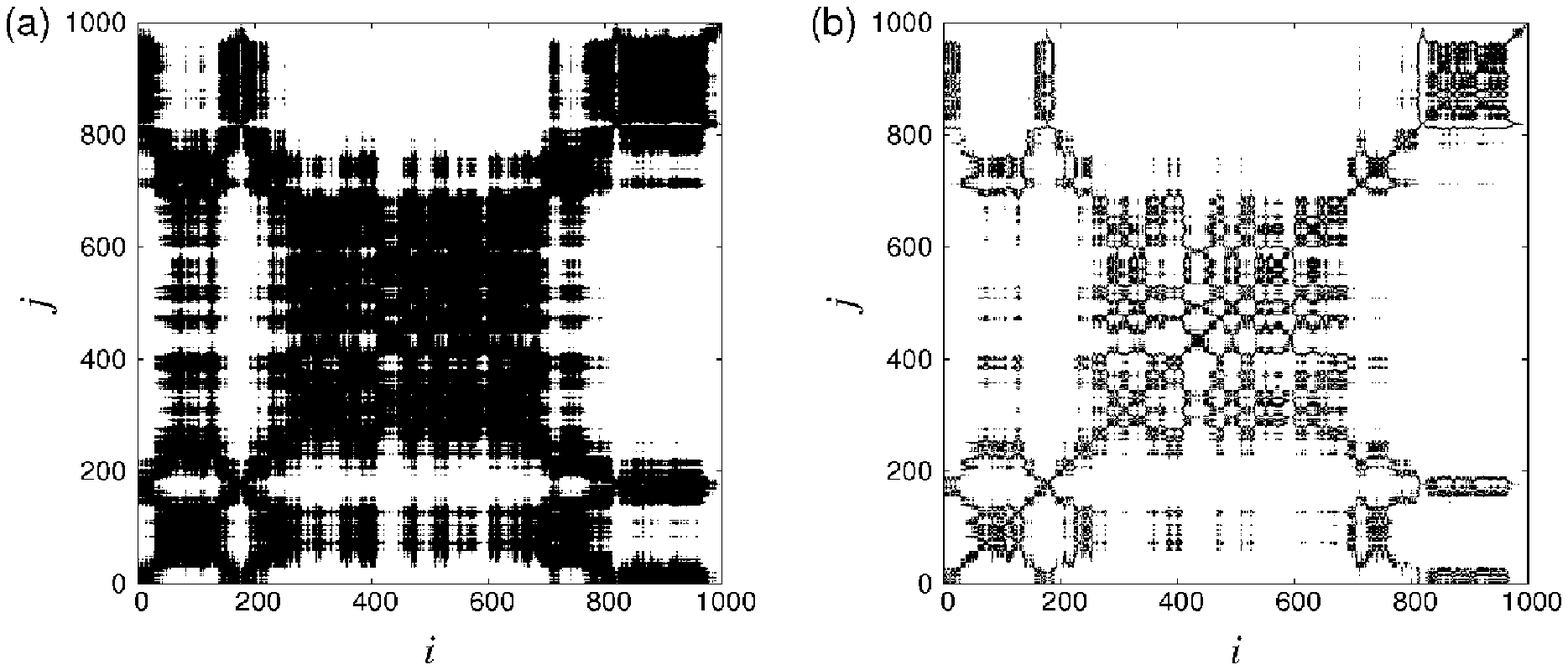}
\caption{\textbf{Recurrence plots for different thresholds}. (a) RP for 1000 days of FTSE 100, $\varepsilon = 10\%$ of maximum distance. (b) RP obtained by normalizing the data in (a), and $\varepsilon = 0.1$, which is around $2\%$ of the maximum distance. This gives a clearer visualization of the finer structure. For both RPs, the time series were not embedded.}
\label{fig7}
\end{figure}
\begin{figure}[tbp]
\centering
\includegraphics[width=\linewidth]{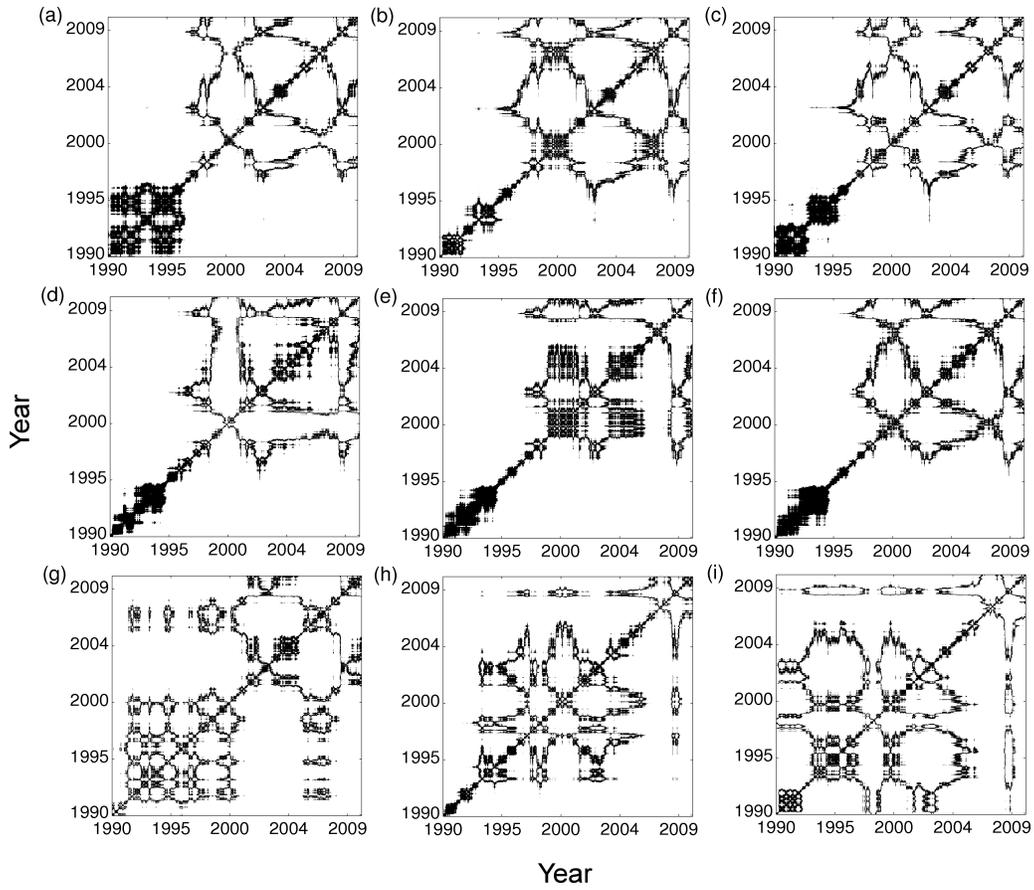}
\caption{\textbf{Recurrence plots for the nine stock indices}. (a) CAC 40. (b) FTSE 100. (c) DAX. (d) NASDAQ. (e) DJIA. (f) S\&P 500. (g) Nikkei 225. (h) Hang Seng. (i) Strait Times. Recurrence threshold $\varepsilon=0.1$; RPs obtained without embedding.}
\label{fig8}
\end{figure}
\begin{figure}[tbp]
\centering
\includegraphics[width=\linewidth]{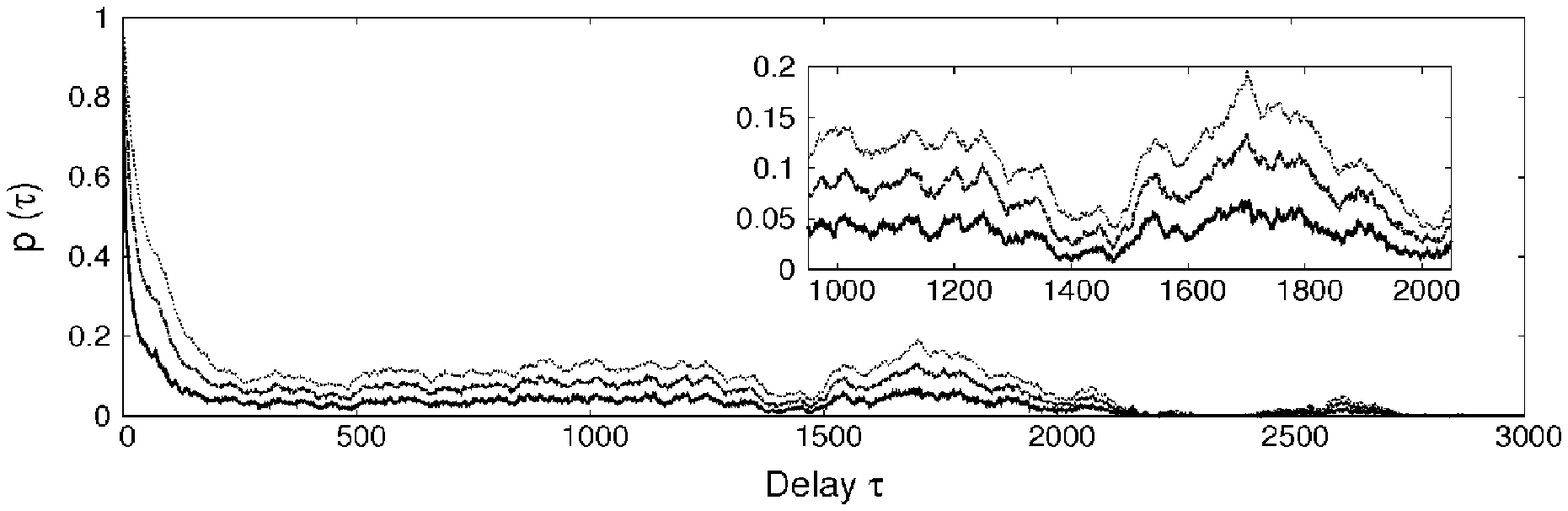}
\caption{\textbf{$p(\tau)$ curves for Hang Seng index}. $\varepsilon = 1\%$ (bottom curve), $\varepsilon = 2\%$ (middle curve), and $\varepsilon = 3\%$ (top curve) of the maximum distance. All three curves have that same pattern. \textit{Inset:} The $p(\tau)$ curves for $\tau = 950$ to $\tau = 2050$ to give a better view of the similar qualitative nature of the three curves.}
\label{fig9}
\end{figure}
In our study we find that keeping $\varepsilon \approx 10\%$ of the maximum distance often results in a `dense' RP (Fig.~\ref{fig7}(a)).	Therefore, to capture the finer recurrence structures while still allowing for sufficient statistics in the $CPR$, we choose $\varepsilon = 0.1$ as the threshold for all RPs and normalize the time series (or any segments thereof) before the recurrence-based calculations. This effectively reduces the recurrence threshold to about $2\%$ of the maximum distance and gives a `clearer' RP (Fig.~\ref{fig7}(b)). The RPs given in Fig.~\ref{fig8} were obtained with $\varepsilon = 0.1$, where it corresponds to around 2-3\% of the maximum distance.\\
We find that the qualitative features of recurrences are robust to this choice of $\varepsilon$. In Fig.~\ref{fig9}, the $p(\tau)$ curves for the Hang Seng index obtained from its RP (Fig.~\ref{fig8}(h)) are shown for $\varepsilon = 1\%,\:2\%$ and $3\%$ of the maximum distance. Although the magnitude of recurrences increases with increasing threshold, the qualitative nature of the curves remains the same throughout. The results obtained in this study are thus robust to small changes in $\varepsilon $.
\subsection{Analyzing connectivity trends: Primary analysis}
\label{connanalysis}
First, a window of 250 time points was moved along pairs of time series and the $CPR$ and $\rho$ values were estimated. This was done for all 36 pairs possible between the nine time series. Next, the $CPR$ values were binned in three categories:
\begin{inparaenum}[(\itshape a\upshape)]
	\item \textit{strong connectedness} ($\lvert \, CPR \, \rvert > 0.8$),
	\item \textit{moderate connectedness} ($0.5 < \lvert \, CPR \, \rvert < 0.8$), and
	\item \textit{weak connectedness} ($\lvert \, CPR \, \rvert < 0.5$),
\end{inparaenum}
	and the number of connections (out of the 36) in each bin were counted for every window.\\
 As a second step, the behavior of $CPR$ during the Dot-Com bubble in 2000 was considered, in which:
\begin{inparaenum}[(\itshape a\upshape)]
	\item taking a pair of time series, the dates when they peaked during the Dot-Com were made to coincide, 
	\item a window of 250 time points was moved from 500 time points before the peak to 250 time points after, and 
	\item the respective $CPR$ (or $\rho$) was obtained.
\end{inparaenum}
\\
At every step, all $CPR$ (or $\rho$) values were tested for significance with 100 Twin Surrogates at 10\% significance.
\section{Results and Discussion}
\label{disc}
In this section, we present the main results obtained using the extended $CPR$ approach and the steps described in Sec.~\ref{connanalysis}.
\subsection{Trends in $CPR$}
\label{cprtrends}
\begin{figure}[tbp]
\centering
\includegraphics[width=\linewidth]{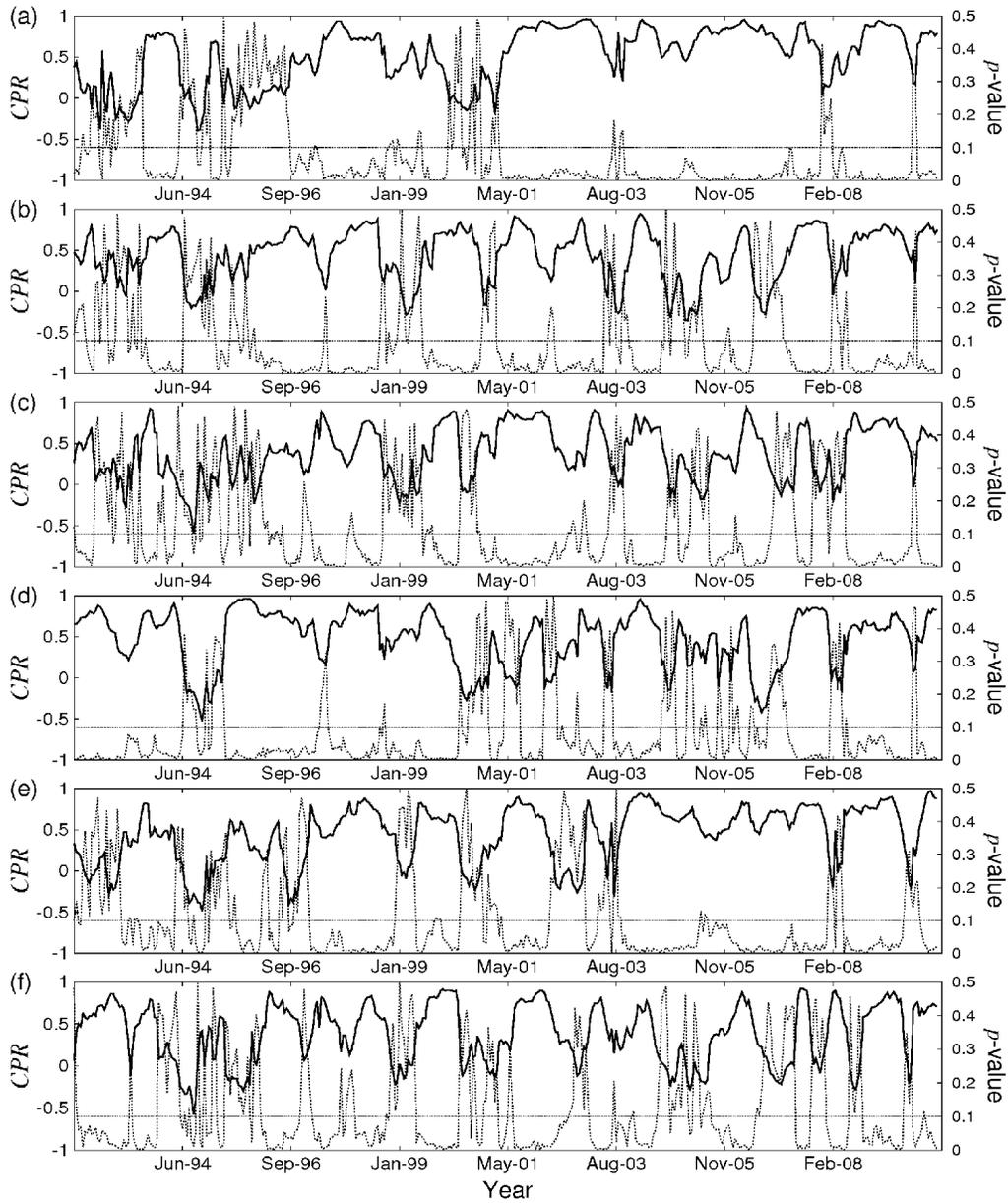}
\caption{\textbf{Trends in $CPR$}. $CPR$ (bold curve) and corresponding $p$-values (dotted curve) for six pairs. (a) CAC 40 and FTSE 100. (b) DAX and NASDAQ. (c) DAX and Nikkei 225. (d) DJIA and S\&P 500. (e) S\&P 500 and Strait Times. (f) Nikkei 225 and Hang Seng. Window size = 250 time points. Step size = 10 time points. The horizontal dotted line denotes the test significance level, $p=0.1$.}
\label{fig10}
\end{figure}
Three points are evident from Fig.~\ref{fig10} (which shows trends in $CPR$ for six pairs of indices along with the corresponding $p$-values, viz.,
\begin{inparaenum}[(\itshape a\upshape)]
	\item CAC 40 and FTSE 100,
	\item DAX and NASDAQ,
	\item DAX and Nikkei 225,
	\item DJIA and S\&P 500,
	\item S\&P 500 and Strait Times, and
	\item Nikkei 225 and Hang Seng).
\end{inparaenum}
\begin{enumerate}[i.]
	\item The $CPR$ does not have a monotonous trend over time. Rather it oscillates erratically with small periods of low $CPR$ interspersing broader bands of high $CPR$.
	\item The regions of low $CPR$ values have high $p$-values (and vice versa), meaning that the lower ranges of $CPR$ tend to be less statistically significant in comparison to the higher values.
	\item These patterns in the $CPR$ are same for all index pairs, of which only six are shown in Fig.~\ref{fig10}. 
\end{enumerate}	
Thus, if $CPR$ were to be interpreted as a measure of similarity, and hence connectedness, it means that the financial world is not moving monotonously to a globally connected scenario, but is rather oscillating between long periods of strong connectedness and short spans of low connectivity. (Nothing conclusive can be said about the nature of connections in the short periods of low $CPR$ as the estimates therein are not statistically significant.) In a sense, what it says is that the more things change the more they remain the same.
\subsection{Patterns of connectivity}
\label{conntrends}
\begin{figure}[tbp]
\centering
\includegraphics[width=\linewidth]{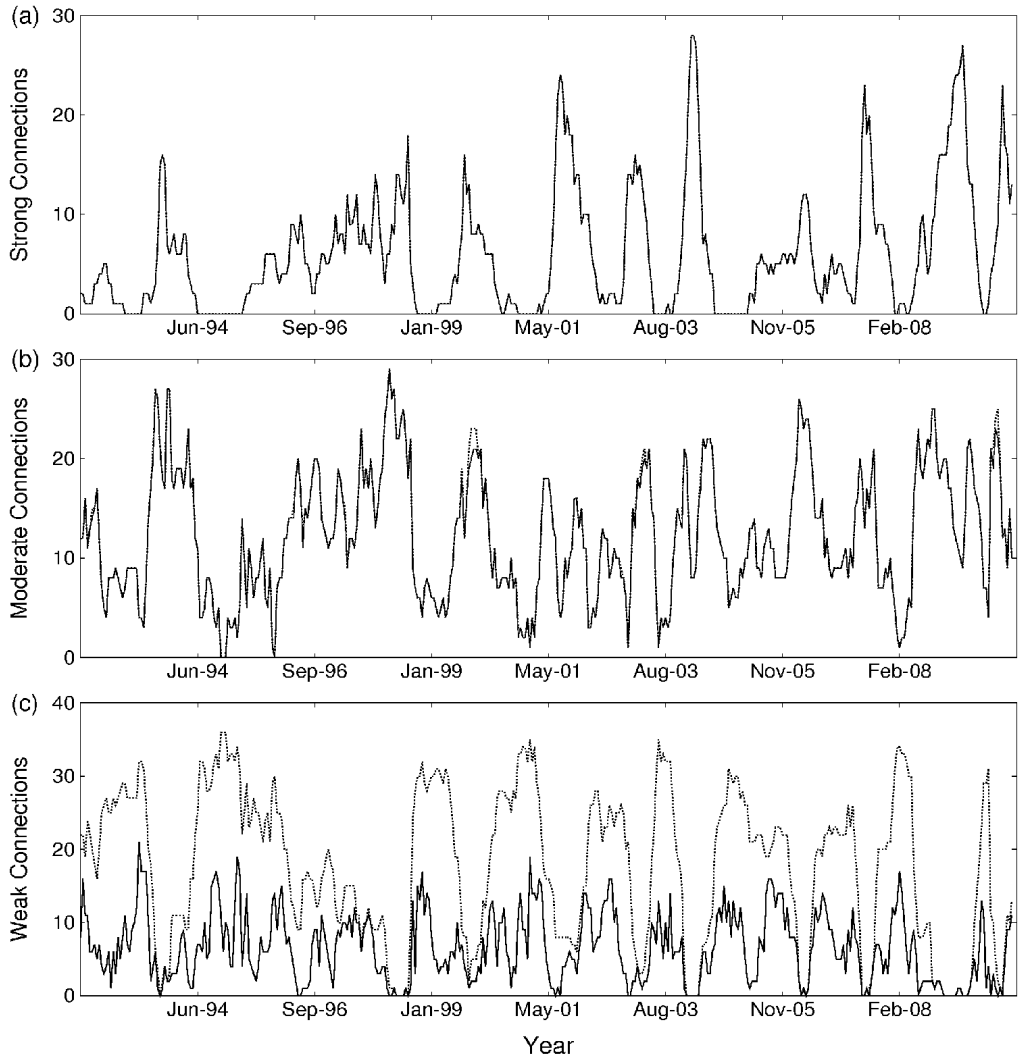}
\caption{\textbf{Patterns of connectivity.} The number of connections in each bin along time. (a) Strong connections. (b) Moderate connections. (c) Weak connections. The bold curves represent the number by counting statistically significant $CPR$ values alone, while the dotted curves represent the number by counting all $CPR$ values in each respective bin. Window size = 250 time points. Step size = 10 time points.}
\label{fig11}
\end{figure}

The patterns in the number of connections in each of the three bins, summed up among all pairs of indices, reinforce the picture put forth in Fig.~\ref{fig10}. This is shown in Fig.~\ref{fig11}. The strong and moderate connections are always statistically significant except for a few instances. The weak connections are clearly more prone to be statistically non-significant. Again, we see that there is no global trend in Fig.~\ref{fig11}, i.e., there is no indication for an increasing global connectivity. Instead, we see that the number of strong and moderate connections oscillate erratically, implying that these nine indices come close to each other, and then move apart, and then come close again, and so on. Had all of them moved to higher connectivity over time, the number of connections in the strong category would have increased progressively, and the numbers in the other two bins would have gone down with it.\\
\begin{figure}[tbp]
\centering
\includegraphics[width=\linewidth]{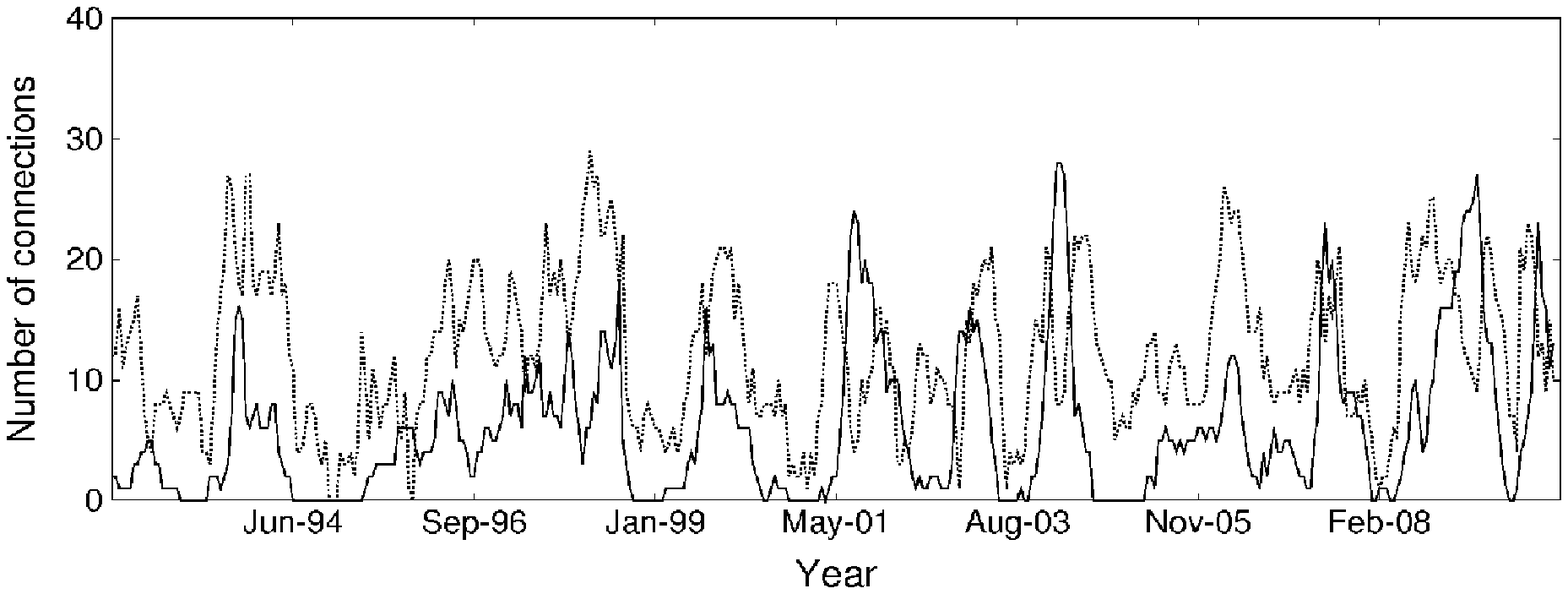}
\caption{\textbf{Strong and moderate connections.} Strong (solid) and moderate (dotted) connections (from Fig.~\ref{fig11} (a) and (b)) shown together. Only statistically significant $CPR$ values were counted.}
\label{fig12}
\end{figure}
The strong and moderate connections move (loosely) in phase with each other (see Fig.~\ref{fig12}). Starting from the dip in Jan-1999, consecutive dips occur roughly at Dec-2000, Jul-2002, Sep-2003, Feb-2005, Sep-2006, Feb-2008, and Oct-2009, meaning that the periods of these dips lie between 14 to 20 months. This might be indicative of the Kitchin business cycle \cite{kitchin23}.\\
Also, the number of `weak' connections (Fig.~\ref{fig11}(c)) is anti-phase to the number of `strong' and `moderate' connections (Figs.~\ref{fig11}(a)-(b) and Fig.~\ref{fig12}), because the total number of connections has to be conserved while the number in each of the three levels of connectedness go through cyclical patterns.
\subsection{$CPR$ in the Dot-Com bubble}
\label{cprdotcom}
\begin{figure}[hp]
\centering
\includegraphics[width=140mm]{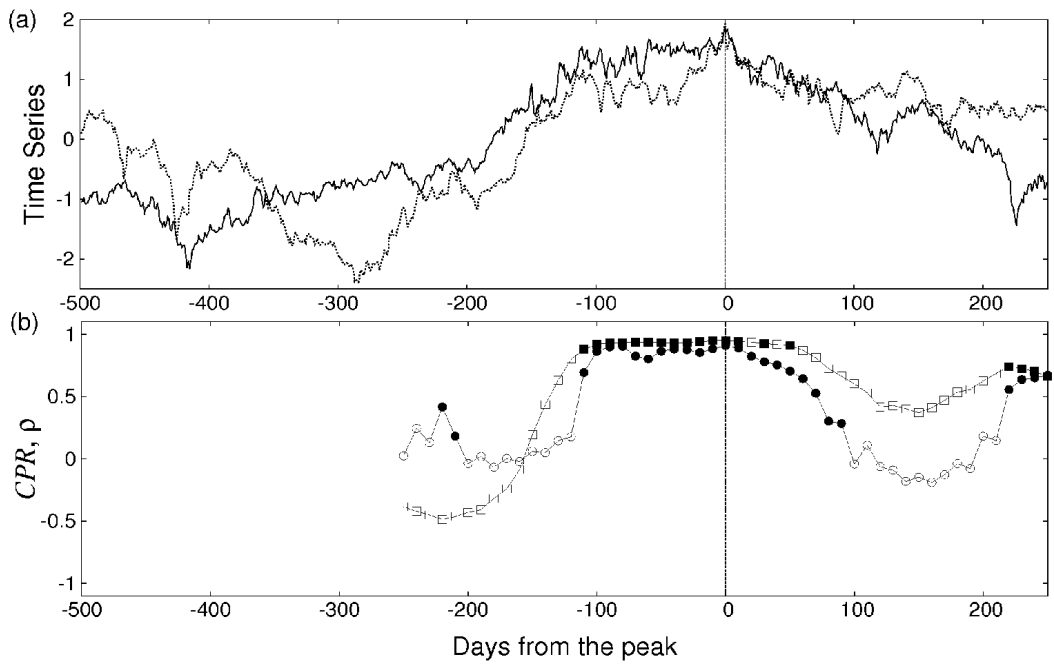}
\caption{\textbf{$CPR$ in the Dot-Com bubble (around the year 2000).} (a) The series CAC 40 (bold) and Strait Times (dotted) with their peak-off date during the bubble coinciding. The vertical dotted line marks the peak-off date. (b) Corresponding $CPR$ (circles) and Pearson's $\rho$ (squares) values. Statistically significant and non-significant values are represented by filled and empty markers respectively.}
\label{fig13}
\end{figure}
\begin{figure}[tbp]
\centering
\includegraphics[width=90mm]{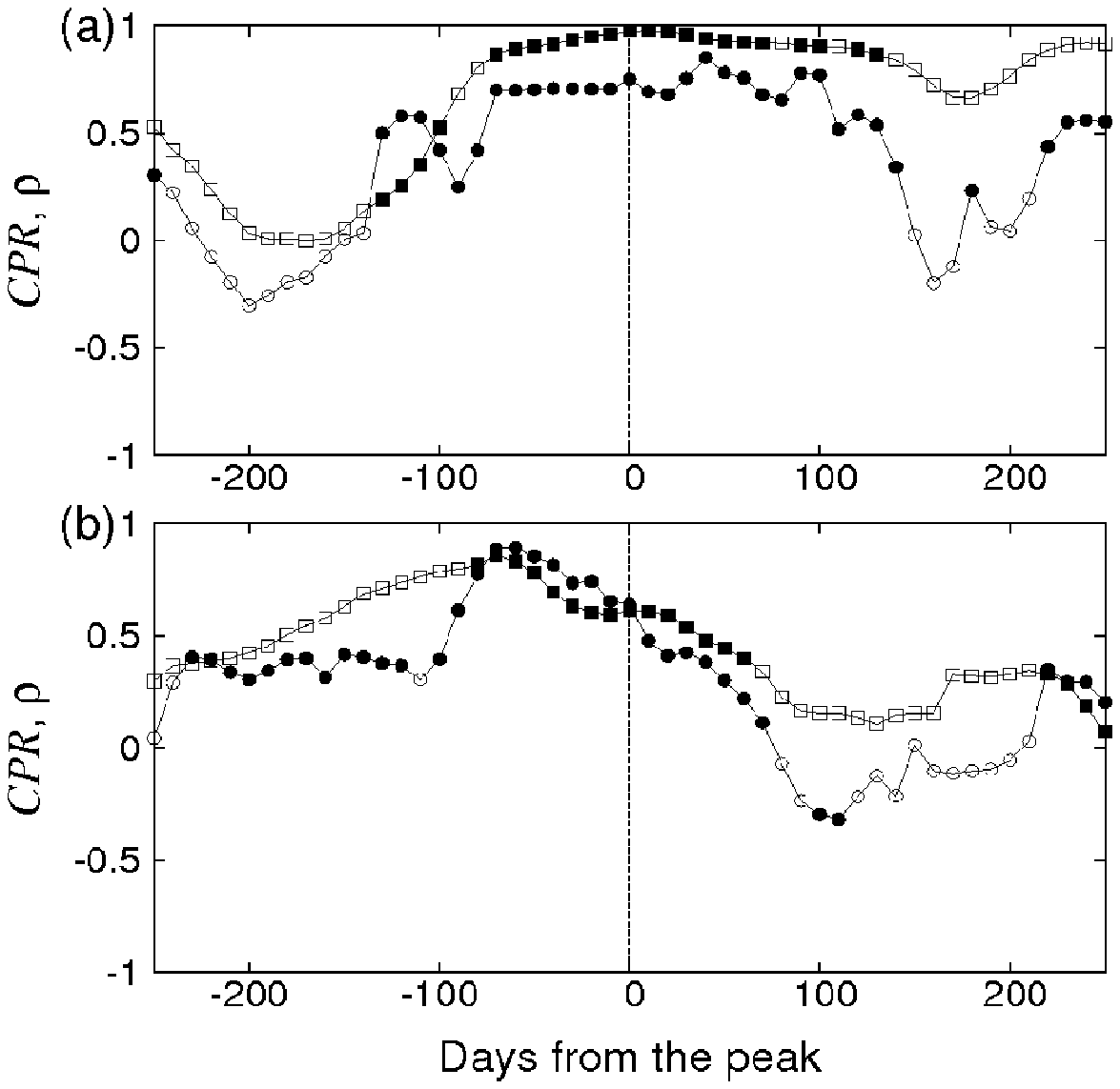}
\caption{\textbf{$CPR$ in the Dot-Com bubble.} (a) DAX vs. NASDAQ. (b) NASDAQ vs. DJIA. Figure keys and markers are the same as in Fig.~\ref{fig13}.}
\label{fig14}
\end{figure}

Figs.~\ref{fig13} and \ref{fig14} give an insight into the way stock indices approach a crisis and then recede from it. Even though the indices peaked and crashed on distinct days (sometimes months apart), once the time series is shifted to align the peak-off dates (Fig.~\ref{fig13}(a)), the $CPR$ increases around the peak-off date and decreases thereafter. For CAC 40 and Strait Times, the decrease in $CPR$ starts almost at the peak-off date (Fig.~\ref{fig13}(b)). On the other hand, for DAX and NASDAQ it occurs after the peak-off date (Fig.~\ref{fig14}(a)) and for NASDAQ and DJIA it happens before (Fig.~\ref{fig14}(b)). The increase-and-decrease of $CPR$ is common to all of them. Thus, the probabilities of recurrence have strong correlation around the peak, i.e., irrespective of the actual date on which a particular index may peak, all indices approach and recede from a crisis similarly.\\
Pearson's $\rho$ also captures this behavior, but $CPR$ is a more sensitive measure, as $\rho$ does not change as sharply as $CPR$ around the peak-off date. Also, the $CPR$ values tend to pass the statistical significance test more often than $\rho$ (see Fig.~\ref{fig13}(b) and Fig.~\ref{fig14}). This point is further elaborated in Sec.~\ref{cpradvantages}.
\subsection{Trends in Pearson's $\rho$}
\label{rhotrends}
\begin{figure}[tbp]
\centering
\includegraphics[width=\linewidth]{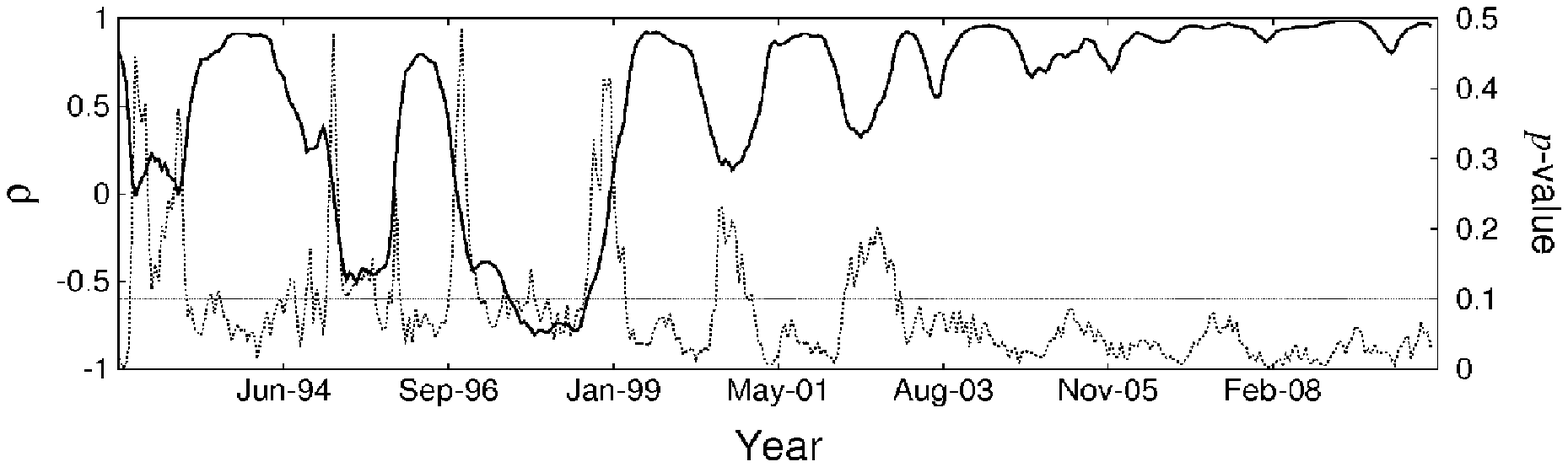}
\caption{\textbf{Trends in Pearson's $\rho$.} Pearson's $\rho$ (bold curve) and its corresponding $p$-values (dotted curve) for the pair S\&P 500 and Strait Times. The horizontal dotted line denotes the test significance level, $p=0.1$.}
\label{fig15}
\end{figure}
The trend in Pearson's $\rho$ is strongly different from $CPR$, as shown for S\&P 500 and Strait Times in Fig.~\ref{fig15} (which is the same pair as in Fig.~\ref{fig10}(e)). Fig.~\ref{fig15} shows an overall movement towards higher correlation as time progresses, in contrast to the oscillating pattern of Fig.~\ref{fig10}(e). The question then arises: \textit{which is the correct picture?} Here, it is crucial to emphasize that these two measures capture different aspects of the time series. While $\rho$ measures the tendency of the time series values to move together in one direction (or opposite directions), $CPR$ measures the tendency of the time series values to return to earlier values at similar time scales. Hence, it is expected that their results are different. However, the case for $CPR$ can be interpreted as follows: being based on recurrence rates it captures the essential dynamical nature of the system. This is in contrast to Pearson's $\rho$ which is simply a statistical comparison of co-evolution of states. Pearson's $\rho$ is less sensitive to changes in the time series as pointed out earlier. Thus, while it may be true that stock indices tend to show more co-movement towards the latter half of our data sets, this does not necessarily imply that they will continue to do so as the corresponding system dynamics move in and out of strong periods of correlation (see Sec.~\ref{cprtrends}).
\subsection{Advantages of $CPR$}
\label{cpradvantages}
\begin{figure}[tbp]
\centering
\includegraphics[width=\linewidth]{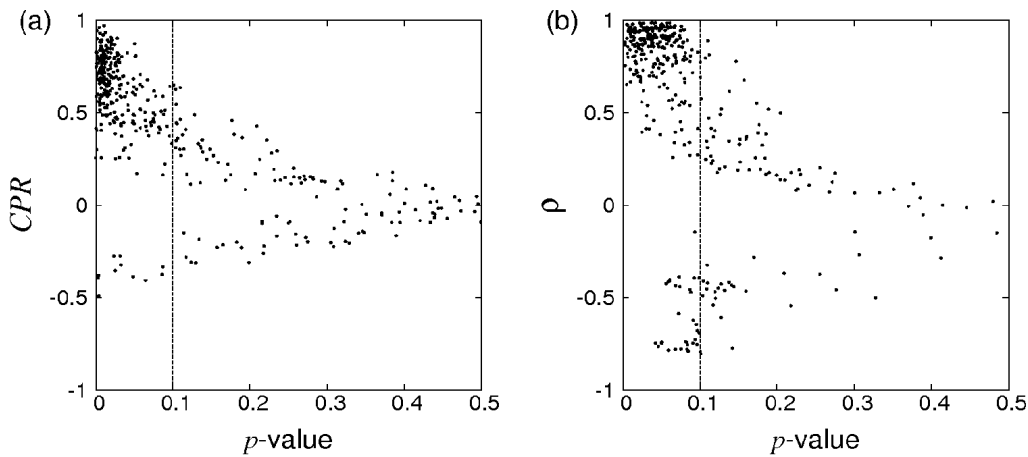}
\caption{\textbf{Patterns of significance.} (a) $CPR$, and (b) Pearson's $\rho$, along with $p$-values.}
\label{fig16}
\end{figure}
We highlight the following advantages of $CPR$ as a measure for estimating connections between financial data sets.
\begin{enumerate}[i.]
	\item It is able to extract patterns even from noisy data sets, as is the case with most financial data. Moreover, it does not require that the data are distributed normally, as is required for the proper usage of $\rho$.
	\item The data sets need not be necessarily embedded for its estimation.
	\item It can be estimated for short time series as well, as is done in the current work.
	\item It does not require high-frequency sampling of the data, which is common in most financial analyses. Our analysis was done with daily sampled data which was freely available on the internet.
	\item It tends to have lower $p$-values than $\rho$, as seen from the spread of points in the top-left corners of the plots in Fig.~\ref{fig16}. Most $CPR$ values are grouped close to the $p=0$ axis, whereas the $\rho$ values have a broader spread. This imples that the probability of rejecting the null hypothesis tends to be lower with $CPR$.
\end{enumerate}
\section{Conclusion}
\label{concl}
In summary, we present a new way of looking at connections between different stock markets. This perspective, involving recurrences and $CPR$, uncovers that, over the past two decades, markets have not proceeded towards an ever increasing connected state but rather moved in and out of strongly connected periods. It also points out that stock indices share similar dynamics during a bubble.\\ Moreover, we highlight the importance of significance testing, with the help of Twin Surrogates, in interpreting measures that analyze field data. In this respect, $CPR$ proves to be a robust measure and moreover, one having the power to reveal patterns from relatively poor data sets --- in terms of noise, low frequency of sampling, and short time series length --- even.\\
Lastly, we suggest a slight modification of the $CPR$, which removes the problem of overestimation of coupling by it, thereby extending its capacity as a measure that can estimate connections between time series data effectively. This opens up newer possibilities of analyzing couplings between financial time series using (cross) recurrence analysis in a better manner.
\section*{Acknowledgments}
B.~Goswami thanks the Deutscher Akademischer Austausch Dienst (DAAD) for funding his internship at the Potsdam Institute for Climate Impact Research (PIK) under the Working Internships in Science and Engineering (WISE) program, wherein this study began. He also thanks Sutirth Dey and Deepak Barua for helpful discussions regarding significance testing and data representation. This work was also partly supported by the Federal Ministry of Education and Research, Germany, (project PROGRESS, 03IS2191B).
\section*{References}
\bibliographystyle{elsarticle-num}
\bibliography{GAmbikaStockRecurrencesRefs}
\end{document}